\documentclass[CPB,twocolumn,10pt,amsmath,superscriptaddress,amssymb,longbibliography]{revtex4}
\usepackage{dcolumn}
\usepackage{amsbsy}
\usepackage{bm}
\usepackage{color,CJK}
\usepackage{amsmath,amssymb,amsfonts}
\usepackage{appendix}
\usepackage{graphicx}
\usepackage{algorithm}
\usepackage{enumerate}
\usepackage{makeidx}
\usepackage{braket}
\usepackage[usenames,dvipsnames]{xcolor}
\usepackage[driverfallback=dvipdfmx,colorlinks,linkcolor=blue,citecolor=blue,urlcolor=blue]{hyperref}

\begin{document}
\title[]
{Production of dual species Bose-Einstein condensates of $^{39}$K and $^{87}$Rb}

\author{Cheng-Dong Mi}
\affiliation{State Key Laboratory of Quantum Optics and Quantum
Optics Devices, \\ Institute of Opto-electronics, Shanxi University,
Taiyuan, Shanxi 030006, People's Republic of China}
\affiliation{Collaborative Innovation Center of Extreme Optics,
Shanxi University, Taiyuan, Shanxi 030006, People's Republic of
China}

\author{Khan Sadiq Nawaz}
\affiliation{State Key Laboratory of Quantum Optics and Quantum
Optics Devices, \\ Institute of Opto-electronics, Shanxi University,
Taiyuan, Shanxi 030006, People's Republic of China}
\affiliation{Collaborative Innovation Center of Extreme Optics,
Shanxi University, Taiyuan, Shanxi 030006, People's Republic of
China}

\author{Peng-Jun Wang}
\email[Corresponding-author:]{ pengjun\underline{ }wang@sxu.edu.cn}
\affiliation{State Key Laboratory of Quantum Optics and Quantum
Optics Devices, \\ Institute of Opto-electronics, Shanxi University,
Taiyuan, Shanxi 030006, People's Republic of China}
\affiliation{Collaborative Innovation Center of Extreme Optics,
Shanxi University, Taiyuan, Shanxi 030006, People's Republic of
China}

\author{Liang-Chao Chen}
\affiliation{State Key Laboratory of Quantum Optics and Quantum
Optics Devices, \\ Institute of Opto-electronics, Shanxi University,
Taiyuan, Shanxi 030006, People's Republic of China}
\affiliation{Collaborative Innovation Center of Extreme Optics,
Shanxi University, Taiyuan, Shanxi 030006, People's Republic of
China}

\author{Zeng-ming Meng}
\affiliation{State Key Laboratory of Quantum Optics and Quantum
Optics Devices, \\ Institute of Opto-electronics, Shanxi University,
Taiyuan, Shanxi 030006, People's Republic of China}
\affiliation{Collaborative Innovation Center of Extreme Optics,
Shanxi University, Taiyuan, Shanxi 030006, People's Republic of
China}

\author{Lianghui Huang}
\affiliation{State Key Laboratory of Quantum Optics and Quantum
Optics Devices, \\ Institute of Opto-electronics, Shanxi University,
Taiyuan, Shanxi 030006, People's Republic of China}
\affiliation{Collaborative Innovation Center of Extreme Optics,
Shanxi University, Taiyuan, Shanxi 030006, People's Republic of
China}

\author{Jing Zhang}
\email[Corresponding author email:]{ jzhang74@sxu.edu.cn,\\ jzhang74@yahoo.com.}
\affiliation{State Key Laboratory of Quantum Optics and Quantum
Optics Devices, \\ Institute of Opto-electronics, Shanxi University,
Taiyuan, Shanxi 030006, People's Republic of China}
\affiliation{Collaborative Innovation Center of Extreme Optics,
Shanxi University, Taiyuan, Shanxi 030006, People's Republic of
China}

\begin{abstract}
We report the production of $^{39}$K and $^{87}$Rb Bose-Einstein
condensates (BECs) in the lowest hyperfine states $| F=1,m_{F}=1
\rangle$ simultaneously. We collect atoms in bright/dark
magneto-optical traps (MOTs) of $^{39}$K/$^{87}$Rb to overcome the
light-assisted losses of $^{39}$K atoms. Gray molasses cooling on
the D1 line of the $^{39}$K is used to effectively increase the
phase density, which improves the loading efficiency of $^{39}$K
into the quadrupole magnetic trap. Simultaneously, the normal
molasses are employed for $^{87}$Rb. After the microwave
evaporation cooling on $^{87}$Rb in the optically plugged magnetic
trap,  the atoms mixture is transferred to a crossed optical
dipole trap, where the collisional properties of the two species
in different combinations of the hyperfine states are studied. The
dual species BECs of $^{39}$K and $^{87}$Rb are obtained by
further evaporative cooling in optical dipole trap at a magnetic
field of 372.6 G with the background repulsive interspecies
scattering length $a_{KRb}$ = 34 $a_{0}$ ($a_{0}$ is the Bohr
radius) and the intraspecies scattering length $a_{K}$ = 20.05
$a_{0}$.
\end{abstract}
\maketitle

\section{Introduction}

Since Bose-Einstein condensates (BECs) were firstly observed in
alkali atoms \cite{bec1, bec2}, the ultracold atomic gases have
attracted much attention to study many interesting quantum
phenomena, e.g. the BEC-BCS crossover, spin orbit coupling
\cite{crossover1,soc1,soc2,soc3}, Mott insulator in the optical
lattice, \cite{wangdw,chen,wang}, and ultracold chemistry
\cite{chemis1,sadiq2}. Dual species ultracold gasses greatly
expand the research area of this platform to study even more
complex physical phenomena that are not accessible in a single
component quantum gas, such as producing heteronuclear molecules
with long dipole-dipole interaction
\cite{molecule1,molecule2,molecule3}, creating polarons near
quantum criticality \cite{polaron1,polaron2,polaron3} and
observation of collective dynamics of a mixture of Bose and Fermi
superfluids \cite{dualsf1}. To date, several dual Bose-Bose
species have been cooled to BECs in experiments, including
$^{23}$Na-$^{87}$Rb \cite{fwang},$^{39}$K-$^{87}$Rb \cite{jarlt},
$^{39}$K-$^{23}$Na \cite{schulze2018}, $^{41}$K-$^{87}$Rb
\cite{modugno, Thalhammer}, $^{88}$Sr-$^{87}$Rb,
$^{84}$Sr-$^{87}$Rb \cite{pasquiou}, Er-Dy \cite{trautmann}, and
$^{133}$Cs-$^{87}$Rb \cite{Lercher, McCarron}.

Recently, more interest is directed towards the $^{39}$K-$^{87}$Rb
dual-species BECs as there is rich Feshbach resonance structure
\cite{jarlt, simoni} for the precise control of the interspecies
scattering length at low magnetic field. Nevertheless, there are
many difficulties in achieving the BEC of $^{39}$K. First, its
negative background sacttering length \cite{chiara} inhibit the
achievement of the BEC state. Second, the unresolved excited-state
hyperfine structure of $^{39}$K \cite{Prevedelli} makes the
standard sub-doppler laser cooling less efficient. Third, the
strong light-assisted losses of the dual species in the initial
double MOTs stage \cite{Marcassa} result in a poor loading. All of
these factors make the achievement of $^{39}$K-$^{87}$Rb
dual-species BECs to be a challenging task.

In this paper, we report the experimental route to overcome these
difficulties, and finally obtain the BECs of $^{39}$K and
$^{87}$Rb in the lowest hyperfine states $| F=1,m_{F}=1 \rangle$
simultaneously. Previously \cite{jarlt}, the $^{39}$K-$^{87}$Rb
BECs were achieved with bright molasses on the D2 line in the $|
F=1,m_{F}=-1 \rangle$ states at around 117.56 G of uniform
magnetic Feshbach resonance field. We follow a slightly different
path to achieve the dual species condensate in the $| F=1,m_{F}=1
\rangle$ states using 372.6 G of uniform magnetic Feshbach
resonance field, and using the gray molasses on the D1 line. At
the three dimensional MOT phases, a dark spontaneous optical force
trap (dark SPOT) \cite{Ketterle} is used to load the $^{87}$Rb
atoms in $| F=1\rangle$ state in the central part without repumping
beams, and a bright MOT loads the $^{39}$K atoms at same time.
This scheme mitigates the atoms loss resulted from the inelastic
collisions and the light-induced loss of different species in the
MOT loading stage. To effectively cool the $^{39}$K, the gray
molasses cooling \cite{dipankar,Salomon,Sievers,shi} on D1 line is
used after the loading stage in magneto-optic trap.
 The obstacle of negative background
sacttering length is overcome by tuning the $^{39}$K scattering
length to $a_{K}$ = 20.05 $a_{0}$ by using an external homogeneous
magnetic field at 372.6 G near a broad Feshbach resonance centered
at 403.4 G, where it is positive to facilitate sympathetic cooling
below 1$\mu$K with the background interspecies scattering length
$a_{KRb}$ = 34 $a_{0}$ during the final evaporation step in the
optical dipole trap. Using this scattering length improves the density of $^{39}$K due to the sympathetic cooling compared to when cooling the $^{39}$K alone.

\section{The $^{39}$K and $^{87}$R$\textbf{b}$ double MOT}

The experimental setup is shown in Fig. 1, which has been used for
the creation of $^{87}$Rb BEC in Ref. \cite{liangchao,guangyu}.
The $^{87}$Rb and $^{39}$K atoms are first simultaneously cooled
and trapped in the 2D MOT chamber to form an atomic beam. This
beam is then pushed to the science chamber (at much lower pressure
of 10$^{-9}$ Pa) with the help of a push beam containing 4
frequencies. In the science chamber, the $^{39}$K atoms are
collected in a 3D MOT while the $^{87}$Rb atoms are collected in a
dark SPOT MOT to reduce the loss of $^{39}$K atoms. We also tried
the double dark SPOT MOT to further reduce the light assisted
losses, however the $^{39}$K dark SPOT MOT did not work.  This
could be due to the use of a single (unbalanced) hollow repumping
beam in contrast to the use of four hollow repumping beams
(balancing each other in the opposite direction) used in
\cite{Stwalley1995}. Since the $^{87}$Rb does not need a high
power repumping beam (compared to the $^{39}$K which needs almost
the same power as that of the trapping/cooling beams due to its
narrowly spaced excited state hyperfine splitting), we do not need
to balance the single hollow repumping beam for the $^{87}$Rb dark
SPOT MOT. The different powers and frequencies used in the push
beam and the two MOT stages are listed in Table I while the Fig. 2
gives the definitions of the detunings.

In the 3D MOT, there are six solid MOT laser beams (2.5 cm
diameter). Each of these six beams carries the Rb and K cooling
and the repumping light only for K, as shown in the insert image
in Fig. 1. To produce the dark region in $^{87}$Rb dark SPOT, an
opaque disk with diameter of 10 mm is used to cast a shadow at the
center of the $^{87}$Rb repumping beam. The center of the hollow
repumping beam is filled with a depumping beam (shown in Fig.
2(b)). The repumping+depumping beam is delivered to the 3D MOT
separately from the six MOT beams and thus the $^{87}$Rb atoms are
trapped at the trap center in $| F=1\rangle$ state out of the
cooling cycle. The depumping beam further depletes the atomic $|
F=2\rangle$ state in the center of the MOT by optical pumping
atoms to the $| F=1\rangle$ state. The dark SPOT MOT significantly
improves the simultaneous loading of both species with reduced
light assisted losses \cite{Marcassa}. The magnetic field gradient
in the 3D MOT stage is 4 G/cm along the $z$-axis which is provided
by a pair of coils in the anti-Helmholtz configuration as shown in
Fig. 1 \cite{fanhao}. The same coils provide the homogeneous
magnetic field for Feshbach resonance when the current direction
in one of the coils is changed.

The setup for MOT lasers is presented in Fig. 3. For $^{87}$Rb,
the cooling and repumping light for the D2 line are generated by
separate lasers due to large ground state hyperfine splitting of
6.8 GHz, and then frequency tuned by double-passed AOMs. The
cooling beam is then amplified by a tapered amplifier (Rb-TA), as
shown in Fig. 3(a). For $^{39}$K, a single master laser can
generate the cooling and repumping lights due to the smaller
ground state hyperfine splitting of 461 MHz. We use two master
lasers, one for the MOT on D2 line and the other one for the gray
molasses on D1 line. The cooling beams from both the master lasers
are first combined together and then amplified by an injection
locked laser. A similar setup amplifies the $^{39}$K repumping
light and all of these are then amplified together by a single
tapered amplifier (K-TA), as shown in Fig. 3(b). This design for
the $^{39}$K guarantees the perfect collimation of the D2 and D1
line laser beams.

The compressed MOT (CMOT) step follows the MOT loading step by
increasing the magnetic field strength to 22 G/cm (duration 150
ms). The fine tuning of the laser parameters during the CMOT
reduces the radiation pressure which together with the high
magnetic field increases the density of the atomic sample. The
repumping+depumping beam for $^{87}$Rb is now blocked and the
repumping light is delivered to the atoms through the six MOT
beams. The detunings and single beam power of the lasers during
the CMOT are shown in Fig. 2(c).

\begin{figure}[h]
\centering
\includegraphics[width=3.3 in]{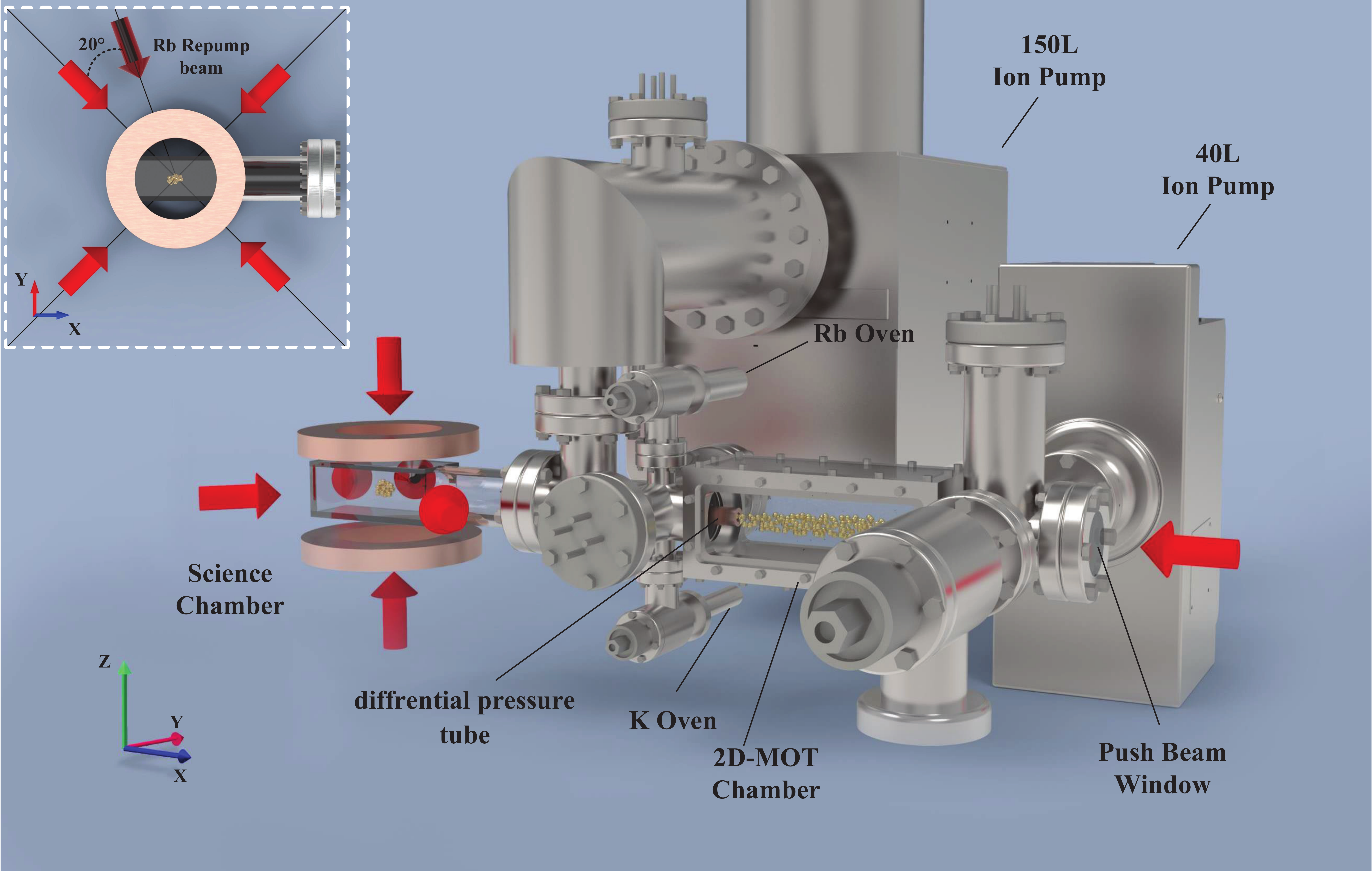}
\caption{  The experimental setup for the dual $^{39}$K and
$^{87}$Rb BECs showing the two MOT cells and the ion pumps for
each chamber. The Rb and K sources (at 40 and 45 $^{0}$C
respectively) supply atoms into the 2D MOT chamber. The science
chamber (enclosed by the quadrupole/Feshbach coils) collects the
atoms in the 3D MOT from the push beams assisted atomic flux
coming from the 2D MOT via a differential pressure tube. The inset
shows the top view of 3D MOT, the four horizontal MOT beams and a
hollow Rb repumping beam are visible.}\label{figure1}
\end{figure}

After the CMOT, the magnetic field is completely turned off in 100
$\mu$s and we perform normal molasses for the $^{87}$Rb but D1
line gray molasses for the $^{39}$K. The D1 line gray molasses
\cite{Salomon} gives samples with higher density and atoms number
compared to the previously devised schemes using the D2 line laser
\cite{jarlt, Landini}. The one-photon detunings $\Delta_{GC}$ and
$\Delta_{GR}$ both are blue detuned +3.33 $\Gamma$ forming a Raman
$\Lambda$-type system in which the dark and bright states formed
by the coherent superposition of the Zeeman sub-levels in the
hyperfine levels $| F=1\rangle$ and $| F=2\rangle$ help cool in
the bright state and hold the atoms in the dark state thus
avoiding the chance of reheating by scattering of photons from the
GM fields. The other powers and detunings are given in Fig. 2(c).
The molasses step lasts for 6.7 ms.

\section{Optical pumping and MW Evaporation}

\begin{figure*}[htbp]
\centering
\includegraphics[width=7 in]{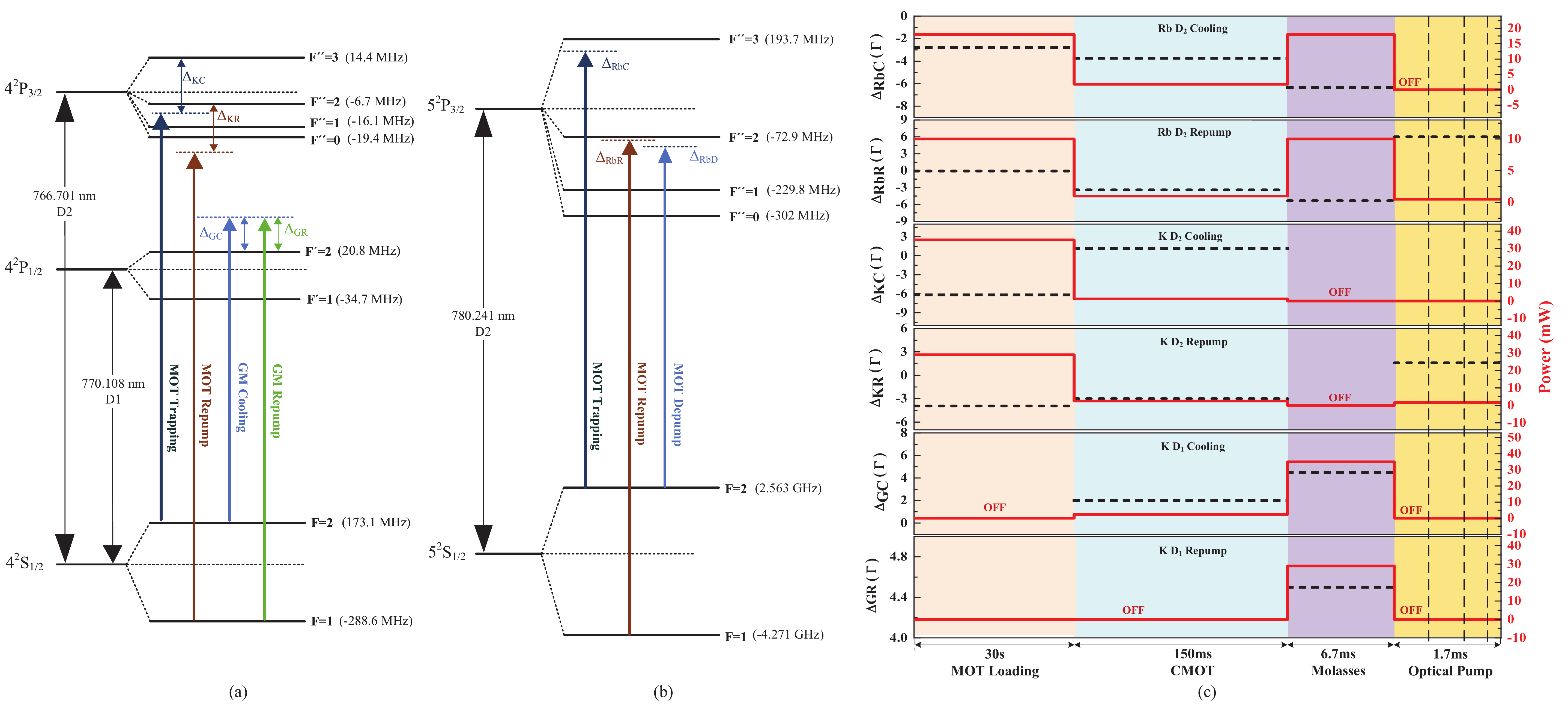}
\caption{ Energy level diagram of $^{39}$K (a) and $^{87}$Rb (b),
and the time sequence of the detunings and powers of each laser.
(a) The $\Delta_{KC}$ and $\Delta_{KR}$ are the detunings of the
trapping and repumping laser from the atomic transitions shown in
the D2 line (766.7 nm) of  $^{39}$K. Similarly $\Delta_{GC}$ and
$\Delta_{GR}$ are the GM cooling and repumping laser detunings
from the transitions shown in the D1 line (770.1 nm). (b)
$\Delta_{RbC}$, $\Delta_{RbR}$ and $\Delta_{RbD}$ are the
detunings of the trapping, repumping and depumping laser
respectively from the transitions shown in the D2 line (780.2 nm)
of $^{87}$Rb. (c) The time sequence of the powers (red solid lines
in the right vertical axis) and detunings (black dashed lines in
the left vertical axis) of the respective lasers is shown. The
duration of each step of the experiment is shown on the horizontal
axis.}\label{figure2}
\end{figure*}

\begin{table}
\centerline{\includegraphics[scale=.5]{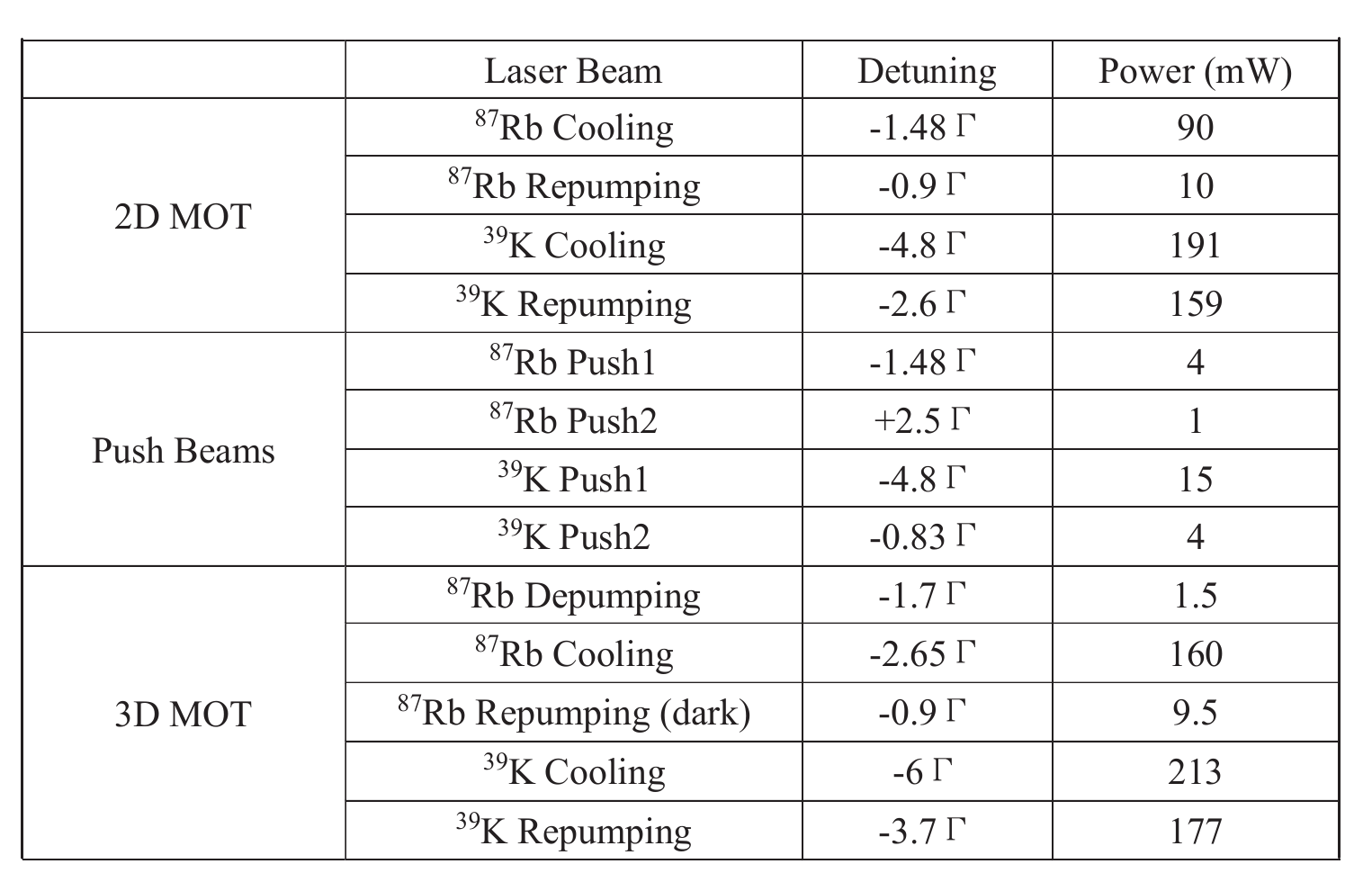}} \vspace{0.06in}

\caption{The detunings and powers of the different laser beams
during the MOT loading. $\Gamma\simeq$6 MHz.
\label{TabI} }

\end{table}

\begin{figure}[htbp]
\centering
\includegraphics[width=3.5 in]{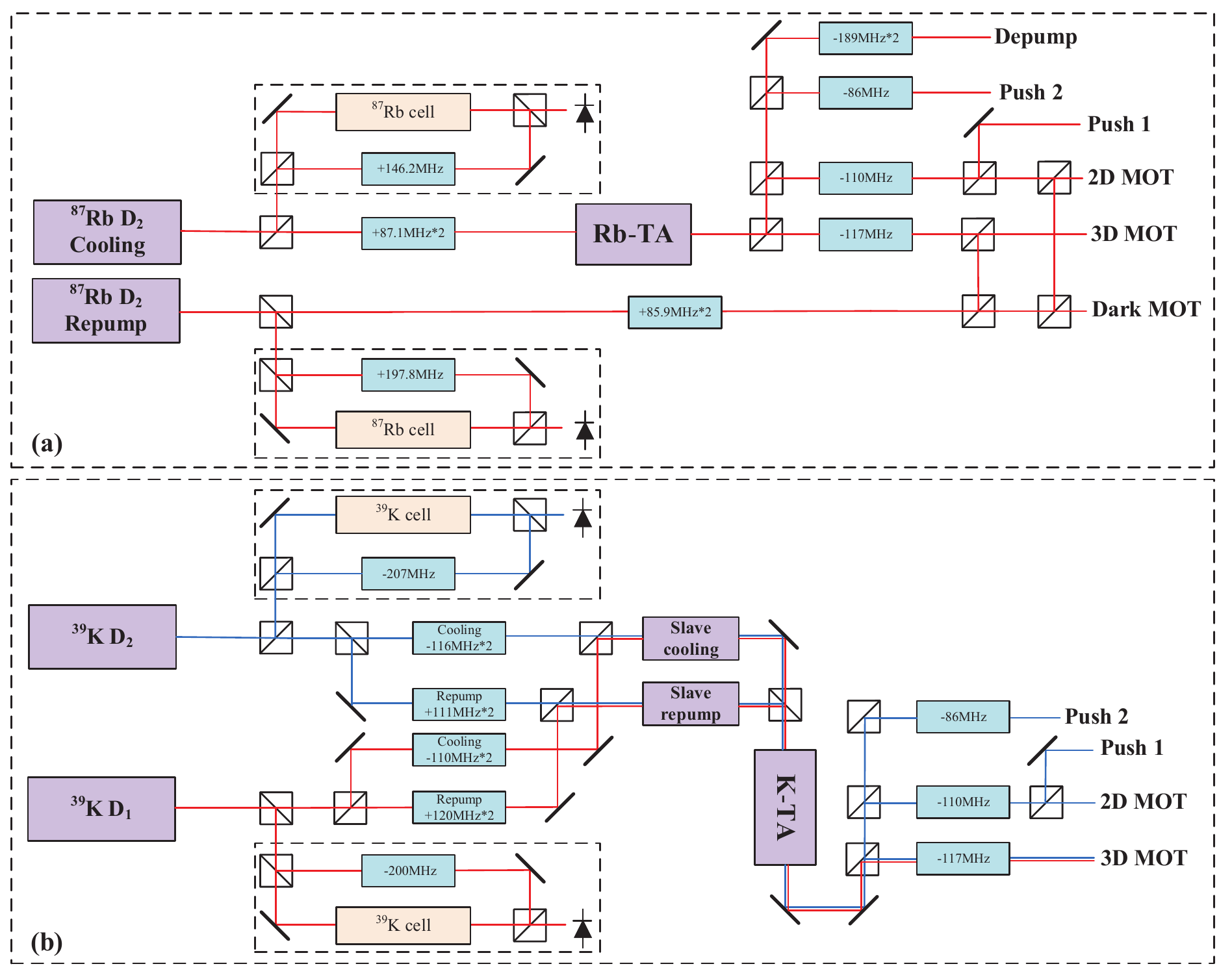}
\caption{ The optical setup and frequency shifting and locking
schemes of the  (a) $^{87}$Rb cooling and repumping lasers and (b)
the $^{39}$K D1 and D2 cooling and repumping lasers. The $^{87}$Rb
repumping laser does not need amplification while all of the
remaining lasers are amplified by tapered amplifiers (TAs) or
injection locked lasers. }\label{figure2}
\end{figure}
Right after the molasses step, the pumping stage follows where
both the $^{39}$K and $^{87}$Rb atoms are optically pumped to the
$| F=2,m_{F}=2 \rangle$  states. This is done by applying a 2 G
homogeneous magnetic field along the $x$-direction and a
$\sigma^{+}$-polarized pump light also along the same direction.
The time sequence for the optimized optical pumping of both the
species is as follows. The repumping light for $^{87}$Rb (detuning
+6 $\Gamma$) remains on in the six MOT beams while the $^{39}$K D2
line repumping light (detuning +1.6 $\Gamma$) is turned on after
the molasses step. Then after 0.5 ms the $^{39}$K D1 line pump
light (on resonance) is turned on for 0.7 ms and the $^{87}$Rb
pump light (4 $\Gamma$) for 1.1 ms. The two repumping beams stay
on for additional 0.1 ms. This completes the optical pumping step.

To efficiently load the sample in the
magnetic quadrupole trap, the magnetic field gradient is ramped up to 26.5 G/cm in 1 ms and held on for
10 ms and then to 62 G/cm in 200 ms and held at this field
for 50 ms. Finally, the magnetic trap is further compressed to 74 G/cm
in 300 ms to increase the atoms density for strong collisions in
the following evaporation stage.  The 20 W green laser is turned on at full power
focused ($\omega_{0}$=30 $\mu$m 1/e$^{2}$ radius) at the center of
the magnetic quadrupole trap to convert it to the optically plugged magnetic quadrupole trap.

Now, the atoms of both species can be cooled by either RF
evaporation or MW evaporation of $^{87}$Rb only \cite{campbell}.
The RF evaporation transfers the hotter $| F=2,m_{F}=2 \rangle$
state atoms to other un-trapped Zeeman states of the same
hyperfine level but the MW evaporation transfers the $|
F=2,m_{F}=2 \rangle$ state atoms to the lower hyperfine
(un-trapped) $| F=1,m_{F}=1 \rangle$ state. The RF evaporation
evaporates both the atomic species but the MW only evaporates the
$^{87}$Rb atoms with sympathetic cooling of the $^{39}$K atoms.
This MW evaporation technique compensates for the low number of
$^{39}$K atoms in the sample at the cost of  $^{87}$Rb atoms.
During the MW evaporation process, the MW frequency is scanned
from 6894.7 to  6855.7 MHz in 6.2 s and then to 6835.7 MHz in
another 5 s. At the end of the evaporation stage, there are around
1.17 $\times 10^{7}$ $^{39}$K atoms and 4.73 $\times 10^{7}$
$^{87}$Rb atoms at about 45 $\mu$K. The magnetic field strength
during the evaporation is held constant at 74 G/cm.

\section{Lifetime Measurement of Different State Mixtures}

After the MW evaporation cooling, the atoms are transferred to the
crossed optical dipole trap, which is made by the intersection of
two 1064 nm laser beams that are frequency shifted (around 10 MHz)
to avoid static interference fringes. Power in both of the dipole
trap beams is ramped up from zero to 3 W in 500 ms.
Simultaneously, the magnetic quadrupole trap strength is ramped
down from 74 G/cm to 7 G/cm in 500 ms thus the atoms are adiabatically transferred to
the optical trap. Three orthogonal pairs of Helmholtz coils are used to
cancel the effect of earth (and any other stray) magnetic fields
on the atoms and hold them stable at the center of the dipole
trap. The dipole trap center is slightly shifted from the magnetic
quadrupole trap center otherwise the green laser plugged in to the
magnetic quadrupole trap center would reduce the efficient loading of
the dipole trap. The green laser is turned off after the loading
ramp is completed. After another 10 ms, the magnetic trap strength is brought down to zero
and the dipole trap alone now holds the atoms. This completes the
optical dipole trap loading.

\begin{figure}[t]
\centering
\includegraphics[width=3.2in]{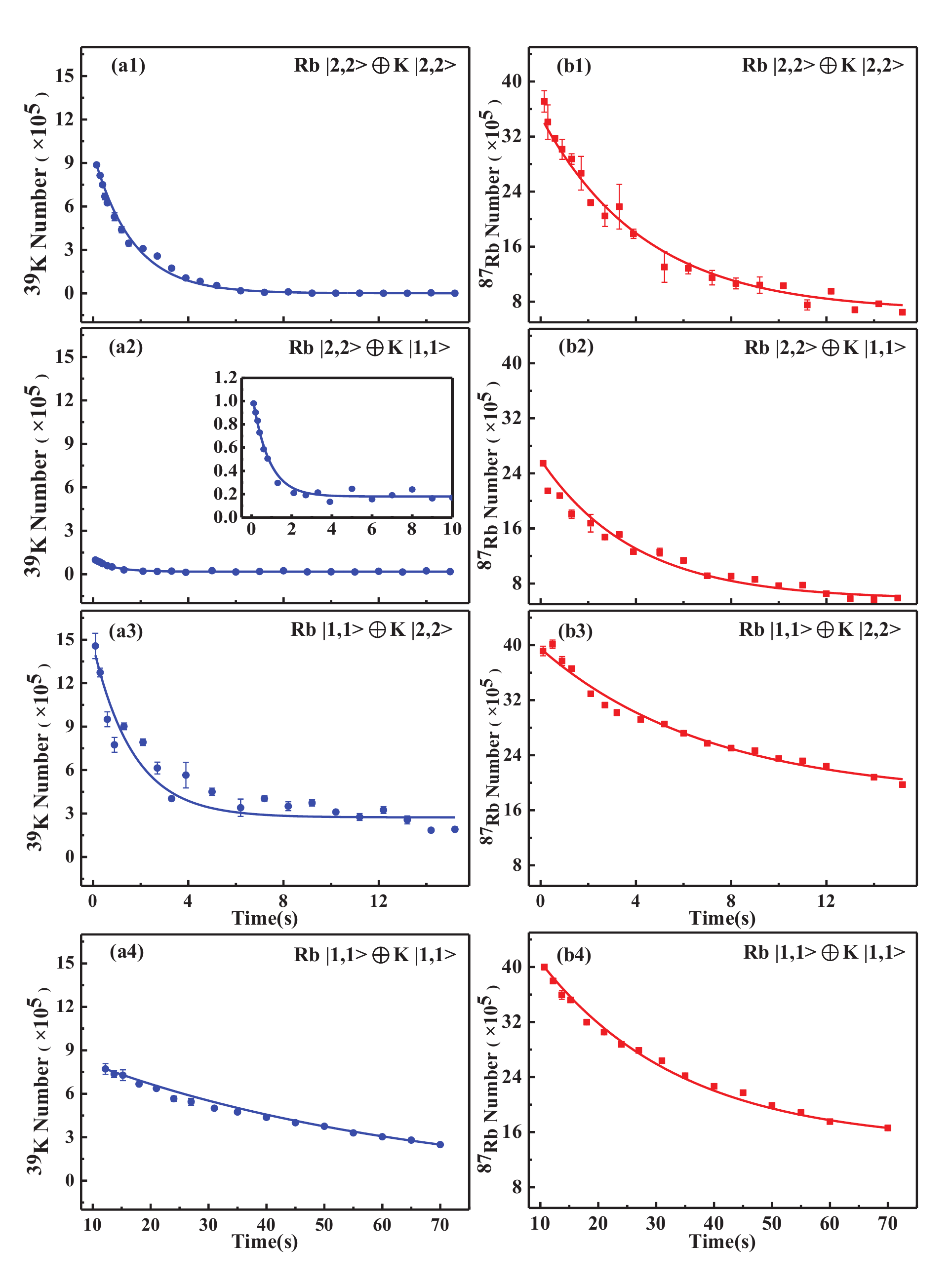}
\caption{ The lifetimes of the various mixture states of $^{39}$K
and $^{87}$Rb. (a) Lifetime of $^{39}$K . (b) Lifetime of
$^{87}$Rb}\label{figure1}
\end{figure}

At this stage, the number of $^{87}$Rb atoms is 8.21
$\times10^{6}$  at 18 $\mu$K while there are 2.15 $\times10^{6}$
$^{39}$K atoms at 21 $\mu$K. The transfer efficiency of the dipole
trap from the quadrupole magnetic trap is around 20 percent for
$^{39}$K and 17 percent for $^{87}$Rb. After the loading is
complete, a background magnetic field is also stabilized at 1 G
along $z$-direction.

A forced evaporation in the optical trap then follows. The dipole
trap power is quickly reduced to 2.6 W in 30 ms to remove the
hotter atoms in the wings of the crossed dipole trap, and then the
power decreases again to 1.6 W in 100 ms. These operations reduce
the temperature of the sample down to 8 $\mu$K. Subsequently, the
$^{87}$Rb atoms are firstly transferred to the $| F=1,m_{F}=1
\rangle$ state adiabatically using a MW sweep of 100 ms duration.
Then a resonant flash light of 1 ms length removes the residual
$^{87}$Rb atoms in the $|2,2 \rangle$ state. After another 1 ms of
wait, the $^{39}$K atoms are also transferred adiabatically to the
$| F=1,m_{F}=1 \rangle$ state using an RF sweep in 100 ms and a
similar flash light follows removing the remaining $^{39}$K atoms
in the $|2,2 \rangle$ state. The transfer efficiency is almost 95
percent. In this spin state preparation process, it is very
important to transfer the $^{87}$Rb atoms to the lower hyperfine
level before $^{39}$K because the ground hyperfine splitting of
$^{87}$Rb is larger than that of $^{39}$K. Therefore, a mixture of
$| F=1,m_{F}=1 \rangle$ state of $^{87}$Rb and $| F=2,m_{F}=2
\rangle$ state of $^{39}$K is more stable than the other way round
which results in extreme losses due to hyperfine changing
collisions \cite{jarlt, Thalhammer, Marcassa}.

To study the loss induced by this collision mechanism, we measure
the lifetime of the different spin state mixtures of the $^{39}$K
and $^{87}$Rb atoms by holding these mixtures in the same dipole
trap depth for variable times and then measuring the number of
remaining atoms. In Fig. 4 the $^{39}$K (a) and $^{87}$Rb (b)
decay curves are shown for four combinations of the hyperfine
states: (1) Rb$|2, 2\rangle$ $\oplus$ K$|2, 2\rangle$, (2) Rb$|2,
2\rangle$ $\oplus$ K$|1, 1\rangle$, (3) Rb$|1, 1\rangle$ $\oplus$
K$|2, 2\rangle$, and (4) Rb$|1, 1\rangle$ $\oplus$ K$|1,
1\rangle$.

The experimental data are fitted by the function
$N_{0}$e$^{-t/\tau}$+$N_{r}$, with $N_{r}$, $N_{0}$ and $t$ as
free parameters ($N_{0}$ is the initial number of atoms and
$N_{r}$ is the atoms remaining after the long holding time), as
the solid lines shown in Fig. 4. The $\tau$ is the lifetime of the
species in a particular mixture. We notice that the mixture of the
two species in case 2 has a very low initial $^{39}$K number
$N_{0}$=0.97$\times10^{5}$, which is one order of magnitude lower
than that in other cases. This is attributed to the faster loss in
the spin state transfer process for $^{39}$K. For case 2, we
obtain a very small lifetime constant of $\tau_{K}$=0.8 s and
$\tau_{Rb}$=3.95 s. We attribute this observed fast decay to the
hyperfine changing collision, which can be described by the
equation below

\begin{equation}\label{eqn}
\begin{split}
& ^{39}K(F=1, m_{F}=1)+^{87}Rb(F=2, m_{F}=2) \rightarrow \\
& ^{39}K(F=2, m_{F}=2)+^{87}Rb(F=1, m_{F}=1)+\Delta,
\end{split}
\end{equation}
where $\Delta$=h$\times$(6334.7-461.7) MHz $\approx$
$k_{B}$$\times$0.306 K. The projection of total hyperfine angular
momentum $m_{F,Rb}+m_{F,K}$ is conserved in such a reaction. The
released energy $\Delta$ in this process converted to the kinetic
energy of the collision atoms, is higher than the trapping
potential of the dipole trap and results in the fast atom loss.
Other hyperfine changing collisions can also happen
\cite{Marcassa} but the main contribution to the losses in our
system is coming from this reaction \cite{simoni,Thalhammer}.
When the spin preparation sequence is reversed (case 3), this type of
hyperfine changing collision is forbidden and longer lifetimes are
observed with $\tau_{K}$=1.7 s and  $\tau_{Rb}$=7.92 s.

\begin{figure}[b]
\centering
\includegraphics[width=3.4 in]{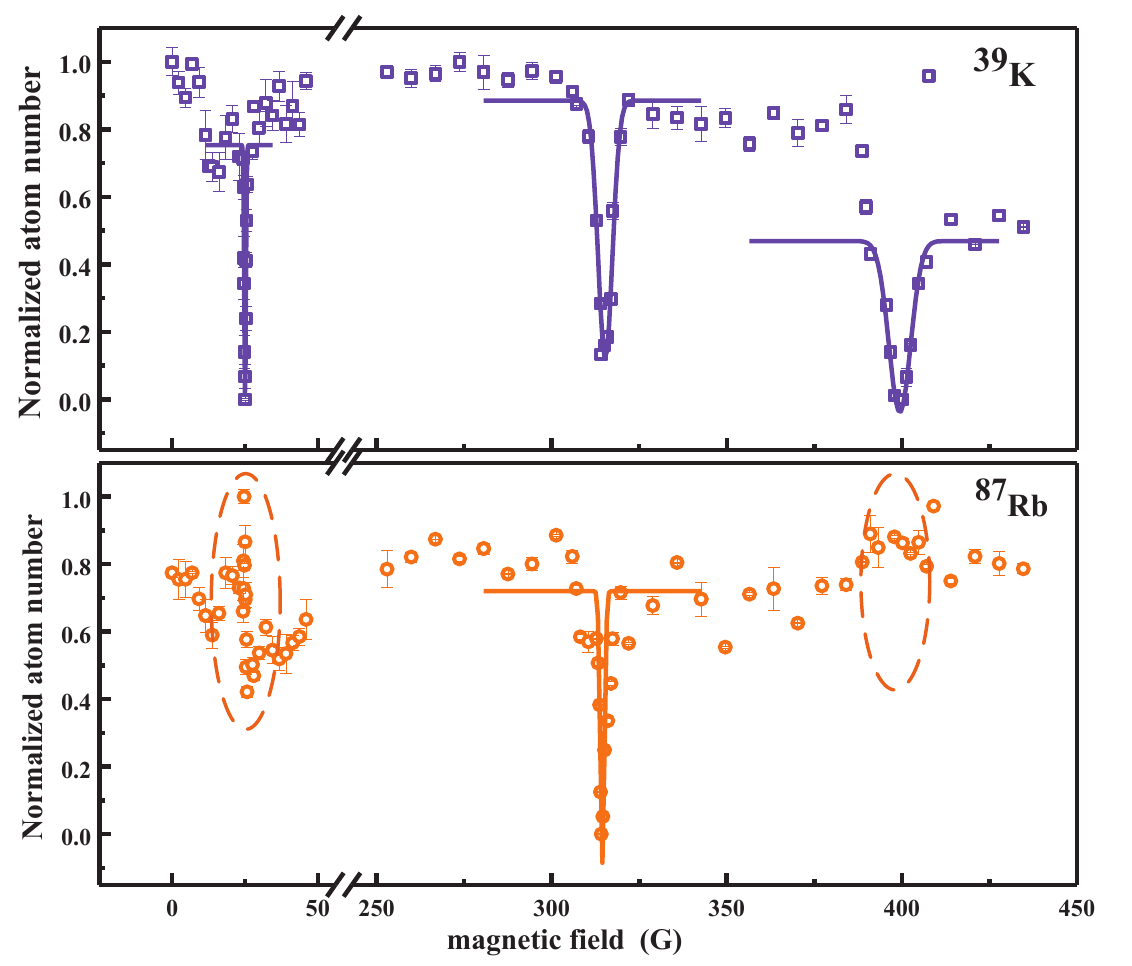}
\caption{ The Feshbach resonance spectrum for the mixture $^{39}$K
and $^{87}$Rb in the $|1,1\rangle$ states. The upper panel shows
the number of atoms remaining in the trap for $^{39}$K and the
lower panel shows that for $^{87}$Rb. The two intra-species
resonances at 25.9 G and 403.4 G for $^{39}$K and the single
inter-species resonance at 318.3 G for the mixture are presnted.}
\label{figure1}
\end{figure}

The measured most stable combination is in the case 4 with longest
lifetimes $\tau_{K}$=28 s and  $\tau_{Rb}$=27 s. Because there is
no any kind of inelastic collision in this combination when both
species are prepared in the lowest states, the lifetime is only
limited by the one-body loss induced by the collision of
background gases and the photon scattering of the dipole trap
beams. The $^{87}$Rb and $^{39}$K mixture when both in the $|2,
2\rangle$ states is not as stable (case 1) as the situation when
each specie is in the $|1, 1\rangle$ state (case 4) because in the
former case there are still some residue atoms in for example the
$|2, 1\rangle$ or other trappable state which can result in
inelastic collisional losses. Losses from other inelastic
collisions reported in \cite{Marcassa} are also possible in this
combination. Therefore, these losses result in $\tau_{K}$=1.65 s
and $\tau_{Rb}$=4.37 s in case 1.

All of these lifetime measurements are done while both of the
species are not condensed (at temperature of 8 $\mu$K). These
decay lifetimes can be of interest for the calculation of the
interspecies scattering lengths among different mixture states at
ultra-cold temperatures.
\section{Feshbach resonances and achievement of dual BEC$\bf{s}$}
  \begin{figure}[b]
\centering
\includegraphics[width=3.4in]{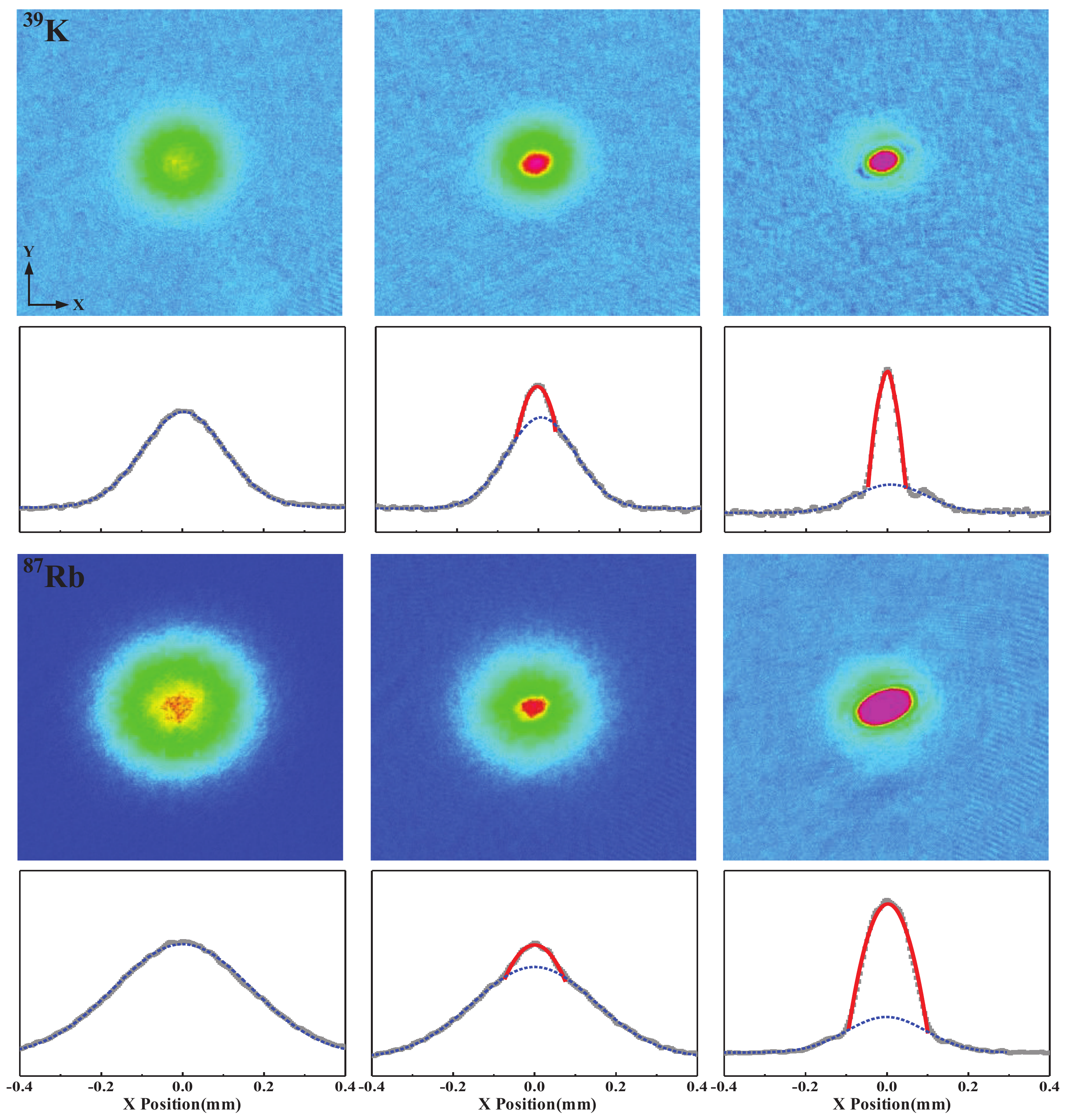}
\caption{ Absorption images of the $^{39}$K (upper row) and
$^{87}$Rb (lower row) atoms near the BEC critical temperature and
the fits to the 1-D integrated optical density along
$x$-direction. The time of flight for these images is 20 ms for
$^{39}$K and 30 ms for $^{87}$Rb. The field of view is 0.8
$\times$ 0.8 mm. }\label{figure1}
\end{figure}
 To further cool the two species, the $^{87}$Rb and $^{39}$K atoms are transferred to the $| F=1,m_{F}=1 \rangle$ state after following the exact
 sequence of state preparation in the dipole trap and then we further perform the evaporation by ramping down the dipole trap power.
 During these ramp down steps, the mixture of $^{39}$K ($a_{K}$ $\sim$ -33$a_{0}$) and $^{87}$Rb ($a_{Rb}$ $\sim$ 100$a_{0}$) atoms (interspecies $a_{RbK}$ $\sim$ 34$a_{0}$) are
cooled by evaporation in the presence of a homogeneous magnetic field of 1 G.
 The first ramp down takes 300 ms and the power in the dipole trap beams is reduced to 0.7 W. In another ramp down lasting for 1 s, the power in the dipole trap beams is reduced to 0.35 W. In the third ramp down, the dipole trap power in the two beams is reduced to 0.26 W in 500 ms.
 At this stage the temperature of the atoms reaches below 1 $\mu$K. At this temperature, we measure the inter- and intra-species Feshbach resonances by scanning the external magnetic field along $z$-direction provided by the pair of coils in the
 Helmholtz configuration (shown in Fig. 1). The resonances are detected as an enhancement of atoms loss \cite{wang2, Gerken, zhangbei}.
 Fig. 5 shows the results of the remaining atom number of both species after holding time of 200 ms for different magnetic field values.

The upper panel shows two $^{39}$K intra-species resonances (25.9 G, 403.4 G) and a
single $^{87}$Rb-$^{39}$K inter-species resonance (318.3 G) recorded by
measuring the number of $^{39}$K atoms after holding the two
species in the magnetic field inside the dipole trap for 200 ms
(including 20 ms ramp up time of the magnetic field). The corresponding number of $^{87}$Rb atoms is shown in the lower panel. The single inter-species resonance is clearly
visible at 318.3 G from the simultaneous loss of both species. The other interesting feature in this spectrum
is the increase in the $^{87}$Rb atoms (encircled) at the exact
$^{39}$K intra-species resonances. This increase in the number of observed $^{87}$Rb atoms is due to the reduced inter-species repulsion caused by the rapid loss of $^{39}$K.

When the atoms mixture is cooled to 0.8 $\mu$K, the negative
background scattering length of $^{39}$K significantly limit the
further effective evaporation cooling. Feshbach resonance can be used
to tune the scattering length to avoid this problem. Many
interesting phenomena can be studied by the tuning of the
scattering length using Feshbach resonances \cite{wang3,
yao1,zhangbei}. To achieve the BECs, the magnetic field of 372.6 G (which
is below the $^{39}$K intra-species resonance at 403.4 G) is
ramped up in 20 ms and held 200 ms for the field stabilization
after the previously mentioned evaporation step of both species in
the $| F=1,m_{F}=1 \rangle$ state. At this time, the scattering
length of $^{39}$K is $a_{K}$ = 20.05$a_{0}$, which is obtained form the well known formula \cite{chiara},

\begin{equation}\label{eqn}
\begin{split}
a(B)=a_{bg}(1-\frac{\Delta}{B-B_{0}}),
\end{split}
\end{equation}
where the background scattering length $a_{bg}$ = -29$a_{0}$, the
Feshbach resonance position $B_{0}$ = 403.4 G and the width
$\Delta$ = -52 G. Then the power in the dipole trap beams is
reduced very slowly from 0.7 W to 0.2 W (in 2000 ms) and the final
dual species BECs are produced.

 To obtain the information of the BECs, the magnetic field is adiabatically ramped to 354.2 G in 30 ms and held for 100 ms for tuning the $^{39}$K scattering length close to zero. Then we perform the time of flight (TOF) imaging. The dipole trap power is completely turned off to let the atom cloud to freely expand for 5 ms in high magnetic field. And then we switch off the high magnetic field and the atoms further freely expand for 20 ms for $^{39}$K and 30 ms for $^{87}$Rb in the background field of 1 G before the absorption imaging. The nearly pure condensates of 4.19 $\times10^{5}$ $^{39}$ K atoms and 5.11 $\times10^{5}$ $^{87}$ Rb atoms are produced.

 Fig. 6 shows the absorption images of the optical density of the $^{39}$K and $^{87}$Rb atoms, the Gaussian fitting for thermal component
 and the polynomial fitting for the condensed fraction. The bimodal character appears in the second columns for both species which
 means a BEC phase transition, and becomes prominent in the third column of Fig. 6 for the almost pure condensates. The first columns of Fig. 6
 for both species show a single mode gaussian fit representing a thermal gas.

\section{Achievement of $^{39}$K BEC by single species evaporation}
\begin{figure}[b]
\centering
\includegraphics[width=3.4in]{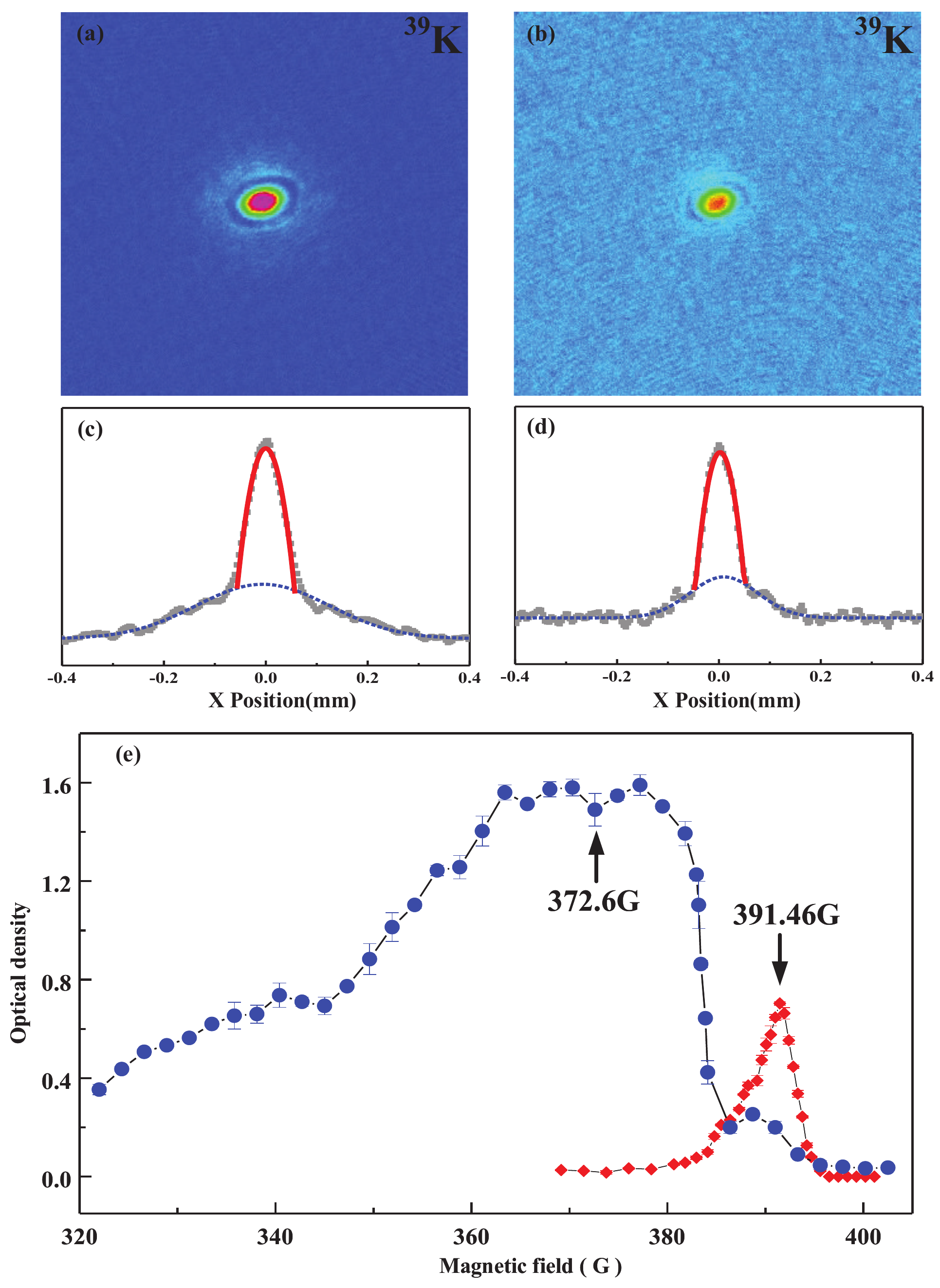}
\caption{ Comparison of the $^{39}$K BEC achieved using the two different
processes. (a) Absorption image of the $^{39}$K BEC achieved using
the two-species evaporation performed at 372.6 G . (b) The same BEC produced using the single-species evaporation performed at 391.46 G. (c, d) The respective poly-line fits to these BECs. (e)
The optical density of the BECs as a function of the magnetic
field in the two processes.}\label{figure1}
\end{figure}

Next, we compare the evaporation efficiency of the two-species and
the single-species of $^{39}$K at different scattering lengths for
$^{39}$K. Here the scattering length $a_{Rb}$ of $^{87}$Rb is
about 100$a_{0}$ and the inter-species scattering $a_{KRb}$ is
about 34$a_{0}$. For the two-species evaporation, we found that a
BEC with high number of atoms can be obtained in the broad region
of positive scattering length $a_{K}$ from about 5.7$a_{0}$ to
35.4$a_{0}$. However, for only  $^{39}$K evaporation the $^{39}$K
BEC can be produced at 97.3$a_{0}$ with low atom numbers as can be
seen from Fig. 7. Here we remove the $^{87}$Rb atoms in the $|2, 2
\rangle$ state using a flash light at 8 $\mu$K after the first two
steps of evaporation in the dipole trap described above.

\section{Conclusion}

In conclusion, we have achieved the dual $^{39}$K and $^{87}$Rb
BECs by utilizing the various available techniques of dark SPOT
MOT, D1 line gray molasses, the microwave evaporation and tuning
the atomic scattering length using the Feshbach resonances. We
also measure the lifetimes of these various mixture states showing
interesting features especially the upper ground stretched states
lifetime is shorter than the lower hyperfine stretched states. The
difference in the lifetimes highlights the importance of the
hyperfine changing collisions in ultra-cold samples of dual
species BECs. These dual BECs with rich Feshbach resonance
structure and large mass imbalance can be used in various
applications such as the formation of heteronuclear quantum
droplets. We also show how the sympathetic cooling of $^{39}$K
using the $^{87}$Rb atoms results in larger and denser BECs
compared to the single species $^{39}$K evaporative cooling.

\section*{Acknowledgments}

This research was supported by the MOST (Grants No.2016YFA0301602
and No.2018YFA0307601), NSFC (Grants No.11974224, No.11704234, No.11804203, No.12034011, No.12022406, No.12004229 and No.92065108),
the Fund for Shanxi ``1331 Project'' Key Subjects Construction, and the
program of Youth Sanjin Scholar.

\end{document}